\RequirePackage{fix-cm}
\documentclass[smallextended]{svjour3}       
\smartqed  
\usepackage{graphicx}
\usepackage{amsmath}
\usepackage{amssymb}
\usepackage{bbm, color}
\begin{document}

\title{Scalable quantum computing model in the circuit-QED lattice with circulator function
}


\author{Mun Dae Kim
\and         Jaewan Kim 
}


\institute{ \at
Korea Institute for Advanced Study, Seoul 02455, Korea \\
              \email{mdkim@kias.re.kr}           
}

\date{Received: date / Accepted: date}

\maketitle

\begin{abstract}
We propose a model for a scalable quantum computing in the
circuit-quantum electrodynamics(QED) architecture. In the Kagome
lattice of qubits three qubits are connected to each other through
a superconducting three-junction flux qubit at the vertices of the
lattice. By controlling one of the three Josephson junction
energies of the intervening flux qubit we can achieve the
circulator function that couples arbitrary pair of two qubits
among three. This selective coupling enables the interaction
between two nearest neighbor qubits in the Kagome lattice, and
further the two-qubit gate operation between any pair of qubits in
the whole lattice by performing consecutive nearest neighbor
two-qubit gates. 
\end{abstract}

\section{Introduction}
\label{intro}

Owing to the remarkable advancements in the qubit (quantum bit) coherence and control
the scalable and programmable quantum computing is expected
to be realized in the near future \cite{Silva,Streltsov,Man,Bromley,Vedral,Brunner,Modi,Haikka,Franco,Orieux,D'Arrigo,Mortezapour,Aaronson}.
A large scale quantum computer consisting of many qubits integrated may
perform quantum algorithms capable of carrying out tasks that are hard or impossible for
ordinary classical computer. Such algorithms can accomplish, for example,
factoring number of large digits \cite{Shor} and searching a large data base \cite{Grover}.
These tasks require the controllability of coupling between two qubits
in the scalable design, which is severely challenging.
We, here, provide an approach to cope with this challenge
by using the  circuit-quantum electrodynamics (QED) architecture \cite{Blais,Blais1}.

In the circuit-QED scheme scalable designs for
one-dimensional array \cite{Nataf,Tian,Leib,Leib2,Viehmann,Kurcz,Raftery}
and two-dimensional lattice system \cite{Houck,Underwood,Schmidt,Peropadre} have been proposed.
There can be two possibilities to control the qubit-qubit interaction
in scalable circuit-QED design.
One is to control the resonator-resonator coupling  by using a intervening
dc-SQUID \cite{Peropadre}, flux qubit \cite{Mariantoni,Reuther,Baust,Wulschner},
Josephson ring modulator \cite{Flurin}, or transmon qubit \cite{bridge}.
On the other hand one can try to tune
the coupling between qubit and transmission line resonator
by controlling the qubit frequency
\cite{Majer,DiCarlo09,DiCarlo,Steffen} or a coupling element inserted between qubit
and resonator \cite{Peropadre2,Romero,Allman,QIP}.


For the universal quantum gate the two-qubit gate
for an arbitrary pair of qubits among three is required.
Hence we need the circulator function which enables  selective coupling
between arbitrary two resonators at the vertex point.
Circulator is a nonreciprocal three-port device that routes a signal
to the next port. Recently Josephson junction based microwave circulators have been proposed
for the quantum information processing with superconducting devices \cite{Koch2,Sliwa}.

In this study we construct the Kagome lattice where three resonators are coupled
at each vertex through an intervening three-Josephson junction flux qubit.
The Josephson junction of the flux qubit consists of dc-SQUID loop
whose effective Josephson junction energy can be controlled
by threading a magnetic flux into the dc-SQUID loop.
By reducing one of three  effective Josephson junction energies
we are able to achieve the microwave circulator function and
to control {\it in situ} the sense of circulation.
We couple qubits to the resonators and then the selective
resonator coupling enables the two-qubit gate between an arbitrary pair of
qubits among three. Further the quantum gate operation between arbitrary pair of qubits
in the whole lattice can be achieved through consecutive two-qubit gates with switching function.

\section{Coupling two circuit-QED cavities}
\label{sec2}

\begin{figure}[b]
\centering
\includegraphics[width=14cm]{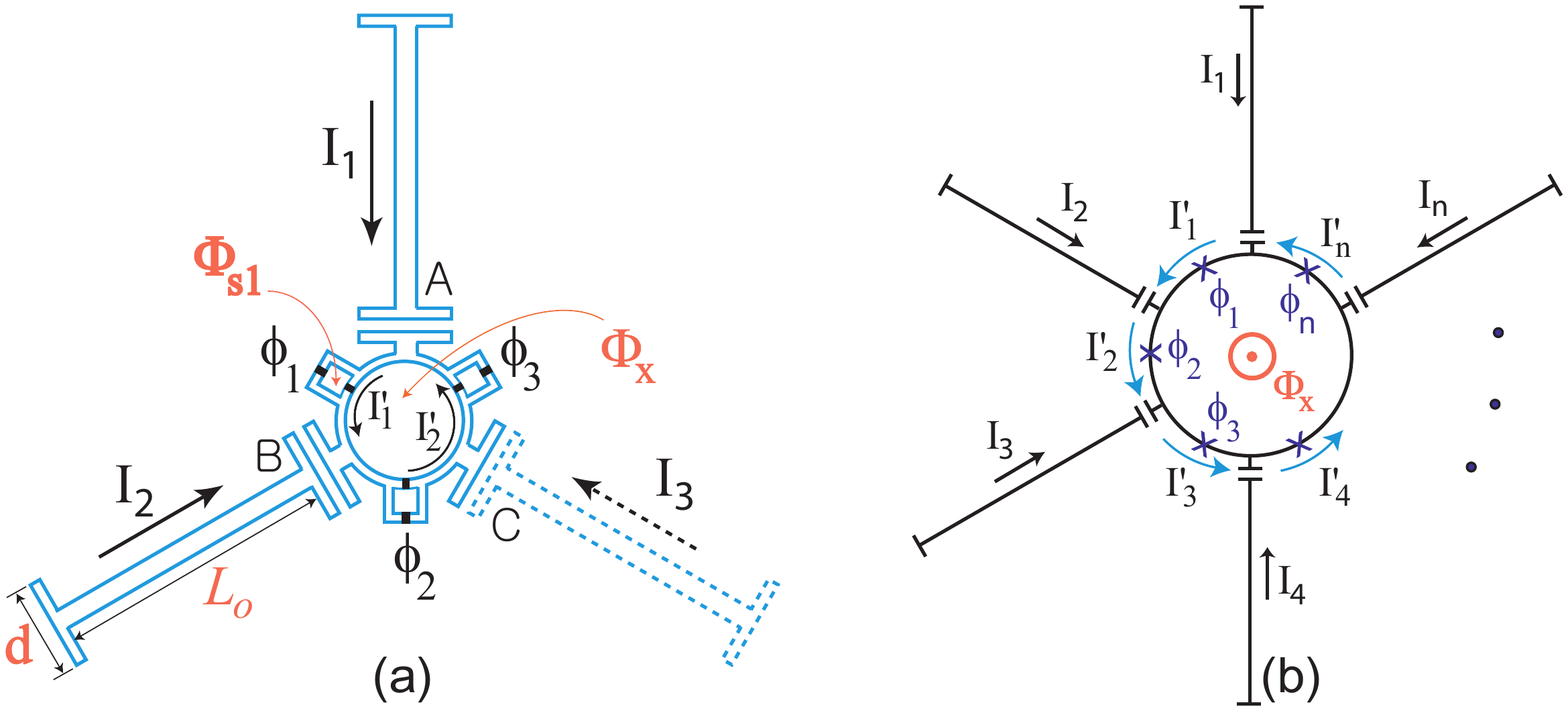}
\vspace{-4cm}
\caption{ (a) Schematic diagram for two resonators
coupled via a three-junction flux qubit through an interface
capacitance at the trisection points A and B. The flux $\Phi_{si}$
threading the dc-SQUID loop controls the effective Josephson coupling
energy of the flux qubit and $\phi_i$ is the phase difference across the dc-SQUID. The
flux threading the three-junction flux qubit loop is usually set as $\Phi_x = 0.5\Phi_0$.
$I_i$ is the current flowing from the resonator to the flux qubit loop through the capacitance,
and $I'_i$ is the current flowing in each segment of flux qubit loop.
(b) n resonators are simultaneously coupled via an n-junction flux qubit loop,
where each dc-SQUID is abbreviated as a single Josephson junction.}
\vspace{-3cm}
\label{fig1}
\end{figure}

First, for simplicity, we consider the case that only two resonators
are coupled through a three-Josephson junction flux qubit
as shown in Fig. \ref{fig1}(a).
There are three
trisection points in the coupling flux qubit, among which two
resonators are connected to the flux qubit at two
points A and B. We will study the case that all three resonators are
coupled to the trisection points A,B and C later.
Here dc-SQUID loops are introduced to
control the effective Josephson coupling energy $E_{Ji}=E_{J}\cos(\pi\Phi_{si}/\Phi_0)$
with the Josephson coupling energy $E_J$ of the junctions in the dc-SQUIDs
and the superconducting unit flux quantum $\Phi_0=h/2e$
by threading an external flux $\Phi_{si}$ into $i$-th dc-SQUID loop.

The periodic boundary condition around the
flux qubit loop \cite{flux,JKPS} becomes
$(1/3)\left(k'_1+2k'_2\right)L'=2\pi(m+f_t)-\phi_1-\phi_2-\phi_3,$
where $m$ is an integer, $k'_i$ the wave vector of the Cooper pair wavefunction
in the flux qubit loop, $\phi_i$ the phase difference across the Josephson junction, $L'$
the circumference of the loop, $f_t=f+f_{\rm ind}$, $f=\Phi_{\rm
x}/\Phi_0$ and $f_{\rm ind}=\Phi_{\rm ind}/\Phi_0$ with the
external magnetic flux $\Phi_{\rm x}$ and the induced magnetic flux
$\Phi_{\rm ind}$.
The induced magnetic flux is given by $\Phi_{\rm ind}=L'_s
(I'_1+2I'_2)/3$, where $L'_s$ is the loop self inductance and
$I'_{i}=-(n_cAq_c/m_c)\hbar k'_{i}$ is the loop current of the flux qubit, and thus we have
$f_{\rm ind}= -(L'_s/L'_K)(L'/3)(k'_1+2k'_2)/2\pi$.
Here, $L'_K = m_cL'/An_cq^2_c$ is the kinetic inductance
\cite{flux}, $A$ the cross section of loop, $q_c=2e$ and $m_c=2m_e$.

Including the induced flux effect the periodic
boundary condition  is written as
\begin{eqnarray}
\label{pbc-1}
\frac{1}{3}\left(1+\frac{L'_s}{L'_k}\right)(k'_1+2k'_2)L'=2\pi(m+f)-\phi_1-\phi_2-\phi_3.
\end{eqnarray}
Further we have the current conservation relations in Fig.
\ref{fig1}(a) such that $I_1=I'_1-I'_2$ and
$I_2=I'_2-I'_1$, that is, $k_1=k'_1-k'_2$ and $k_2=k'_2-k'_1$,
and thus $k'_1-k'_2=(1/2)(k_1-k_2)$, where $I_{i}=-(n_cAq_c/m_c)\hbar k_{i}$.
On the other hand the current relation,
$I'_i=-I_{ci}\sin\phi_i+C'_i{\dot V}_i$, of the
capacitively-shunted model of Josephson junction can be represented as
\begin{eqnarray} \label{Jc}
-(n_cAq_c/m_c)\hbar k'_i=-I_{{\rm
c}i}\sin\phi_{i}-C'_{i}(\Phi_0/2\pi)\ddot{\phi}_{i}
\end{eqnarray}
by using  the Josephson voltage-current relation
$V_i=-(\Phi_0/2\pi){\dot \phi}_i$ with $C'_i$ being the capacitance
of  Josephson junction,  $n_c$  the Cooper pair density and
$I_{{\rm c}i}=2\pi E_{Ji}/\Phi_0$ the critical current of
Josephson junction.

From these relations we can obtain the equation of motion for the
phase variable $\phi_j$,
\begin{eqnarray}
\label{motion}
\left(\frac{\Phi_0}{2\pi}\right)^2C'_{j}\ddot{\phi}_{j}&=&
\frac{\Phi^2_0}{2(L'_s+L'_K)\pi}\left(m+f-\frac{1}{2\pi}\sum^3_{i=1}\phi_i\right)\nonumber\\
&&-E_{Jj}\sin\phi_{j}+p_j\frac{\Phi_0}{12\pi}(I_1-I_2),
\end{eqnarray}
where $p_1=-2$ and $p_2=p_3=1$. This equation of motion can be
derived from the Lagrange equation, $(d/dt)\partial{\cal
L}/\partial{\dot \phi}_i-\partial{\cal L}/\partial \phi_i=0$, with
the Lagrangian
\begin{eqnarray}
\label{Lag} {\cal L}(\phi_i,\dot{\phi}_i)&=&\sum^3_{i=1}\frac12
C'_i\left(\frac{\Phi_0}{2\pi}\right)^2\dot{\phi}^2_i
-U_{\rm eff}(\{\phi_i\}),\\
\label{Ueff}
U_{\rm eff}(\{\phi_i\})&=&\frac{\Phi^2_0}{2(L'_s+L'_K)}\left(m+f-\frac{1}{2\pi}\sum^3_{i=1}\phi_i\right)^2
+\sum^3_{i=1}E_{Ji}(1-\cos\phi_i)\nonumber\\
&&-\frac{\Phi_0}{12\pi}(I_1-I_2)(\phi_2+\phi_3-2\phi_1).
\end{eqnarray}
Hence we can consider this Lagrangian describes the dynamics of the system in Fig. \ref{fig1}(a).

For the usual parameter regime for the
three-Josephson junction flux qubit \cite{Chiorescu}
we can neglect $L'_K$ because $L'_K/L'_s\sim 0.01$.
Further, the inductive energy $\Phi^2_0/2L'_s$ dominates over the other energy scales
so that  the first term in Eq. (\ref{Ueff})
can be removed leaving the usual flux quantization condition,
$2\pi(m+f)-\sum^3_{i=1}\phi_i\approx 0$ at the minimum energy ~\cite{Orlando}.
By using this constraint with $m=0$ and introducing the coordinate
$\phi_\pm=(\phi_2\pm\phi_3)/2$ the effective potential can be
transformed to
\begin{eqnarray}
U_{\rm eff}(\phi_+,\phi_-)&\approx& -E_{J1}\cos(2\pi f-2\phi_+)-2E_J\cos\phi_+\cos\phi_-\nonumber\\
&&-\frac{\Phi_0}{6\pi}(I_1-I_2)(-2\pi f+3\phi_+), \label{Ueff+-}
\end{eqnarray}
where we consider that one junction has smaller Josephson coupling
energy while two junctions larger one such that $E_{J1} <E_{J2}=E_{J3}=E_J$.

\begin{figure}[b]
\vspace{-3cm} \centering
\includegraphics[width=8cm]{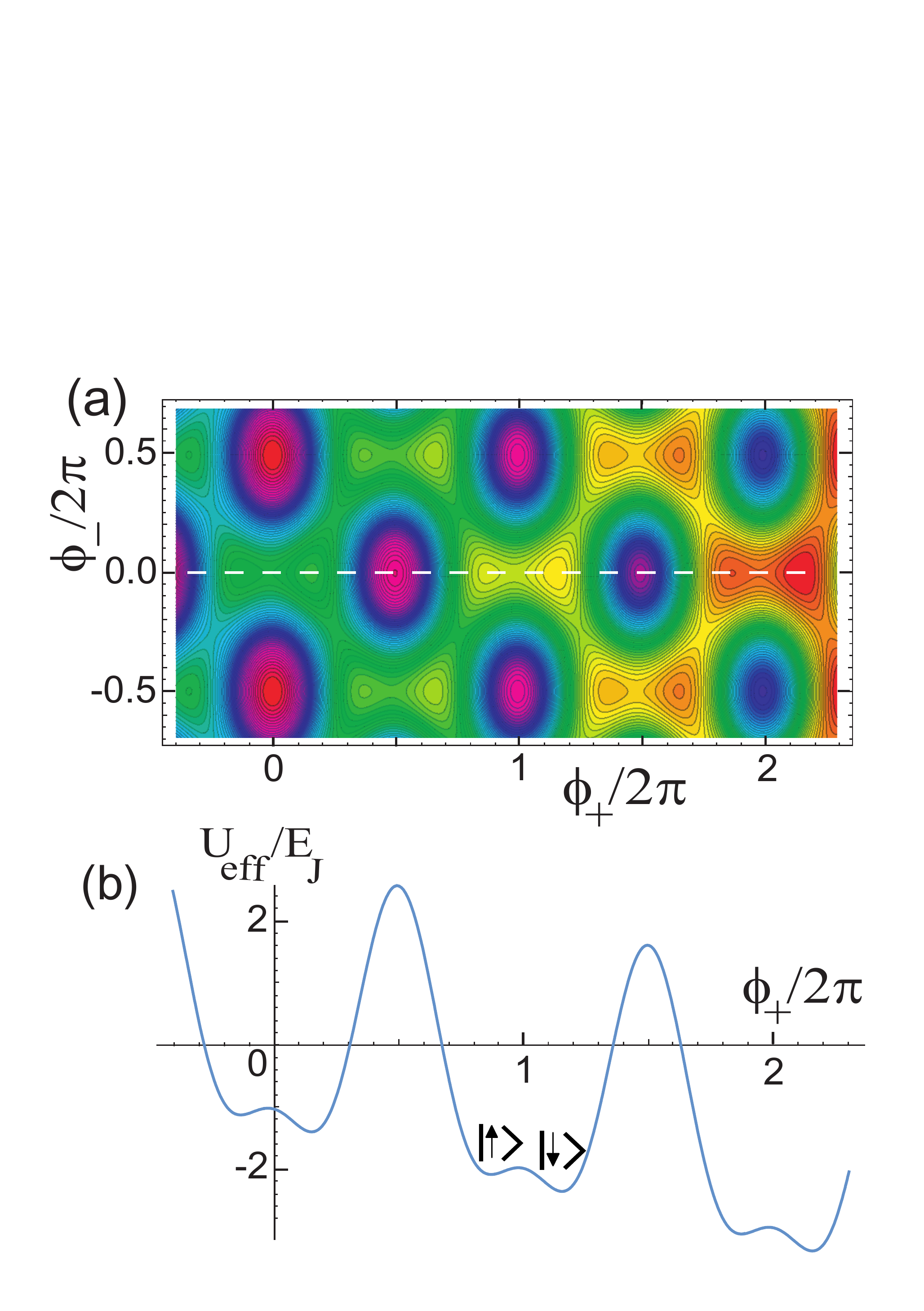}
\vspace{-0cm}
\caption{(a) Effective potential $U_{\rm
eff}(\phi_+,\phi_-)$ of the coupled two-resonator system in Fig.
\ref{fig1}(a). We set $(I_1-I_2)/I_c=0.05, f=0.5$,
and $E_{J1}/E_J=0.8$.(b) Potential profile along the dashed line
in (a). The resonator current $I_1-I_2$ tilts the potential and
thus the degeneracy of the current states,
$|\uparrow\rangle$ and $|\downarrow\rangle$,  is broken.}
\label{fig2}
\end{figure}

In Fig. \ref{fig2}(a) we show the effective potential
$U_{\rm eff}(\phi_+,\phi_-)$. If there is no current from resonators,
$I_1=I_2=0$, the effective potential $U_{\rm eff}(\phi_+,\phi_-)$
has local minima along $\phi_-=0$ and $\phi_+=\alpha (-\alpha)$
with $\alpha=|\phi_{\rm +,min}|$ at the local minima,
corresponding to the counterclockwise (clockwise) current state,
$|\downarrow\rangle$ ($|\uparrow\rangle$), of the flux qubit loop. In the tight-binding
approximation the Hamiltonian can be written as $H=E_\downarrow
|\downarrow\rangle\langle\downarrow|+
E_\uparrow|\uparrow\rangle\langle\uparrow|
-t_q(|\downarrow\rangle\langle\uparrow|+|\uparrow\rangle\langle\downarrow|)$
with $t_q$ being the tunneling amplitude between the potential
local minima when two states, $|\downarrow\rangle$ and $|\uparrow\rangle$,
are degenerate, $E_\downarrow=E_\uparrow$. If a finite dc-bias current $I_1-I_2$
is applied, the effective potential becomes tilted, whose profile
along the dashed line in Fig. \ref{fig2}(a) is
shown in Fig. \ref{fig2}(b). As a result, one of the
two energy levels increases while the other decreases, and thus
the degeneracy of the diagonal terms of the Hamiltonian $H$ become
broken such as
$(E_\downarrow-\Phi_0\alpha(I_1-I_2)/2\pi)|\downarrow\rangle\langle\downarrow|
+(E_\uparrow+\Phi_0\alpha(I_1-I_2)/2\pi)|\uparrow\rangle\langle\uparrow|$.
Since $E_{J1} < E_{J2}=E_{J3}=E_J$ and the junction with larger
Josephson coupling energy has a smaller phase difference, we have
$\alpha\lesssim \pi/3$ and $\phi_2=\phi_3, {\it i. e.}, \phi_-=0$
at the potential minima in the central part of Fig. \ref{fig2}(b).

Consider two resonators are coupled to the intervening flux qubit loop at the
end of the resonators  as shown in Fig. \ref{fig1}(a).
In the design we introduce a large capacitance between two line segments of length $d$
through which the current flows between a resonator and the intervening qubit loop.
In this study we consider the second harmonic mode of resonator for later purpose.
The second harmonic mode of current in $j$-th resonator can be represented as
${\cal I}_j(x,t)=-i\sqrt{\hbar\omega_{rj}/l_sL}\sin(2\pi
x/L)[a_j(t)-a^\dagger_j(t)]$  ~\cite{Blais,JKPS} in terms of the boson operator
$a_j(t)$ with $\omega_{rj}$ being the frequency of
resonator mode, $l_s$ the inductance density and $L=L_0+d$ the effective
length of resonator. Here, the origin of x-coordinate is at the center of resonator.

The charge fluctuation in the resonator induces the current
flowing into the flux qubit loop given by
$I_j(t)=\int^{L/2}_{L_0/2}{\dot q}_j(x,t)dx$ which can be
represented by the difference of currents at both ends of the
interface, $I_j(t)={\cal I}_j(L/2,t)-{\cal I}_j(L_0/2,t)$ by using
the current conservation ${\dot q}_j(x,t)=\partial {\cal
I}_j(x,t)/\partial x$. Hence the current flowing from the $j$-th
resonator can be written as
\begin{eqnarray}
\label{I0}
I_j(t)
=-i\sqrt{\frac{\hbar\omega_{rj}}{l_sL}}\sin\left(\frac{\pi d}{L}\right)[a_j(t)-a^\dagger_j(t)].
\end{eqnarray}

We can put this ac-current of resonator into the effective
potential of Eq. (\ref{Ueff+-}),
and then the total Hamiltonian is represented in the basis of
$|0\rangle=(|\downarrow\rangle+|\uparrow\rangle)/\sqrt{2}$ and
$|1\rangle=(|\downarrow\rangle-|\uparrow\rangle)/\sqrt{2}$ as
follows:
\begin{eqnarray}
\label{Hc} {\cal H}&=&\hbar\sum_{j=1,2}\omega_{rj} a^\dagger_j
a_j+\frac{1}{2}\omega_a\sigma_z+\frac{i}{2}
g\sigma_x[(a_1-a^\dagger_1)-(a_2-a^\dagger_2)],
\end{eqnarray}
with the coupling strength $g\approx \alpha \Phi_0\sqrt{\hbar\omega_{r}/l_sL}(d/L)$
and $\omega_a=2t_q$, where we set $\omega_{rj}=\omega_r$.
This Hamiltonian  describes the interaction between the resonator
modes 1 and 2. The last term of Hamiltonian ${\cal H}$ shows that
a photon in the resonator 1 excites the flux qubit state and then the
flux qubit goes back to the ground state, generating a photon in
resonator 2, and vice versa. Therefore, two resonators are coupled
by using the flux qubit as an intervening qubit mediating the
interaction \cite{bridge}.
%

\section{Two-qubit gate in the Kagome lattice of qubits}
\label{sec3}

Now we consider the case that all three trisection points $A, B$,
and $C$ are coupled to resonators  in Fig.
\ref{fig1}(a). Generally $n$ resonators can be coupled to
the intervening flux qubit as shown in Fig.
\ref{fig1}(b). Then the periodic boundary condition
around the flux qubit loop is given by
$(L'/n)\sum^n_{i=1}k'_i=2\pi(m+f_t)-\sum^n_{i=1}\phi_i$. From the
relations $\Phi_{\rm ind}=L'_s\sum^n_{i=1}I'_i/n$ and
$I_{i}=-(n_cAq_c/m_c)\hbar k_{i}$, we have   $f_{\rm
ind}=-(L'_s/L'_K)(L'/n)\sum^n_{i=1}k'_i/2\pi$, and thus
\begin{eqnarray}
\left(1+\frac{L'_s}{L'_K}\right)\sum^n_{i=1}k'_i\frac{L'}{n}
=2\pi(m+f)-\sum^n_{i=1}\phi_i.
\end{eqnarray}

By evaluating $k'_i$ from this periodic boundary condition and the
current conservation  $I'_1=I'_n+I_1, I'_2=I'_1+I_2,
I'_3=I'_2+I_3, \cdots , I'_n=I'_{n-1}+I_n$, that is,
$k'_1=k'_n+k_1, k'_2=k'_1+k_2, k'_3=k'_2+k_3, \cdots ,
k'_n=k'_{n-1}+k_n$, and putting $k'_i$ into the current relation
of capacitively-shunted model of Josephson junction,
$I'_i=-I_{ci}\sin\phi_i+C'_i{\dot V}_i$, we obtain an equation of
motion for $\phi_j$,
\begin{eqnarray}
\label{3motion}
\left(\frac{\Phi_0}{2\pi}\right)^2C'_j\ddot{\phi}_j&=&\frac{\Phi^2_0}{2(L'_s+L'_K)\pi}
\left(m+f-\frac{1}{2\pi}\sum^n_{i=1}\phi_i\right)\nonumber\\
&&-E_{Jj}\sin\phi_{j}+\frac{\Phi_0}{2\pi n}\sum^n_{i=1}((n-i+j)
~{\rm mod} ~n)I_i.
\end{eqnarray}
This equation of motion can be obtained  from the Lagrange equation with the
effective potential
\begin{eqnarray}
\label{Ueffn} U_{\rm eff}(\{\phi_i\})&=&
\frac{\Phi^2_0}{2L'_s}\left(m+f-\frac{1}{2\pi}\sum^n_{i=1}\phi_i\right)^2
+\sum^n_{i=1}E_{Ji}(1-\cos\phi_i)\\
&&-\frac{\Phi_0}{2\pi n}\sum^n_{i,j=1}((n-i+j) ~{\rm mod} ~n)\phi_j I_i. \nonumber
\end{eqnarray}

For $n=3$, specifically, the last term of the effective potential $U_{\rm
eff}(\{\phi_i\})$ can be represented as $(\Phi_0/6\pi) [\phi_1(I_2-I_3)+\phi_2(I_3-I_1)+\phi_3(I_1-I_2)].$
In Fig. \ref{fig1}(b) we
consider that the junction with phase difference $\phi_1$ has smaller
Josephson coupling energy while two junctions larger one such that
$E_{J1} < E_{J2}=E_{J3}=E_J$.
Then, from the analysis similar to that in
the previous section with the boundary condition
$\sum^3_{i=1}\phi_i-2\pi(m+f)=0$ the effective potential $U_{\rm
eff}(\{\phi_i\})$ in Eq. (\ref{Ueffn}) for $n=3$ can be rewritten
in the transformed coordinate $\phi_\pm=(\phi_2\pm\phi_3)/2$ as
\begin{eqnarray}
\label{UeffT3}
U_{\rm eff}(\phi_+,\phi_-)&=&-E_{J1}\cos(2\pi
f-2\phi_+) -2E_{J}\cos\phi_+\cos\phi_- \nonumber\\
&&-\frac{\Phi_0}{6\pi}[(\!-\!2\pi f\!+\!3\phi_+)(I_1\!-\!I_2)\!+\!\phi_-(2I_3\!-\!I_1\!-\!I_2)].
\end{eqnarray}

\begin{figure}[b]
\centering
\includegraphics[width=10cm]{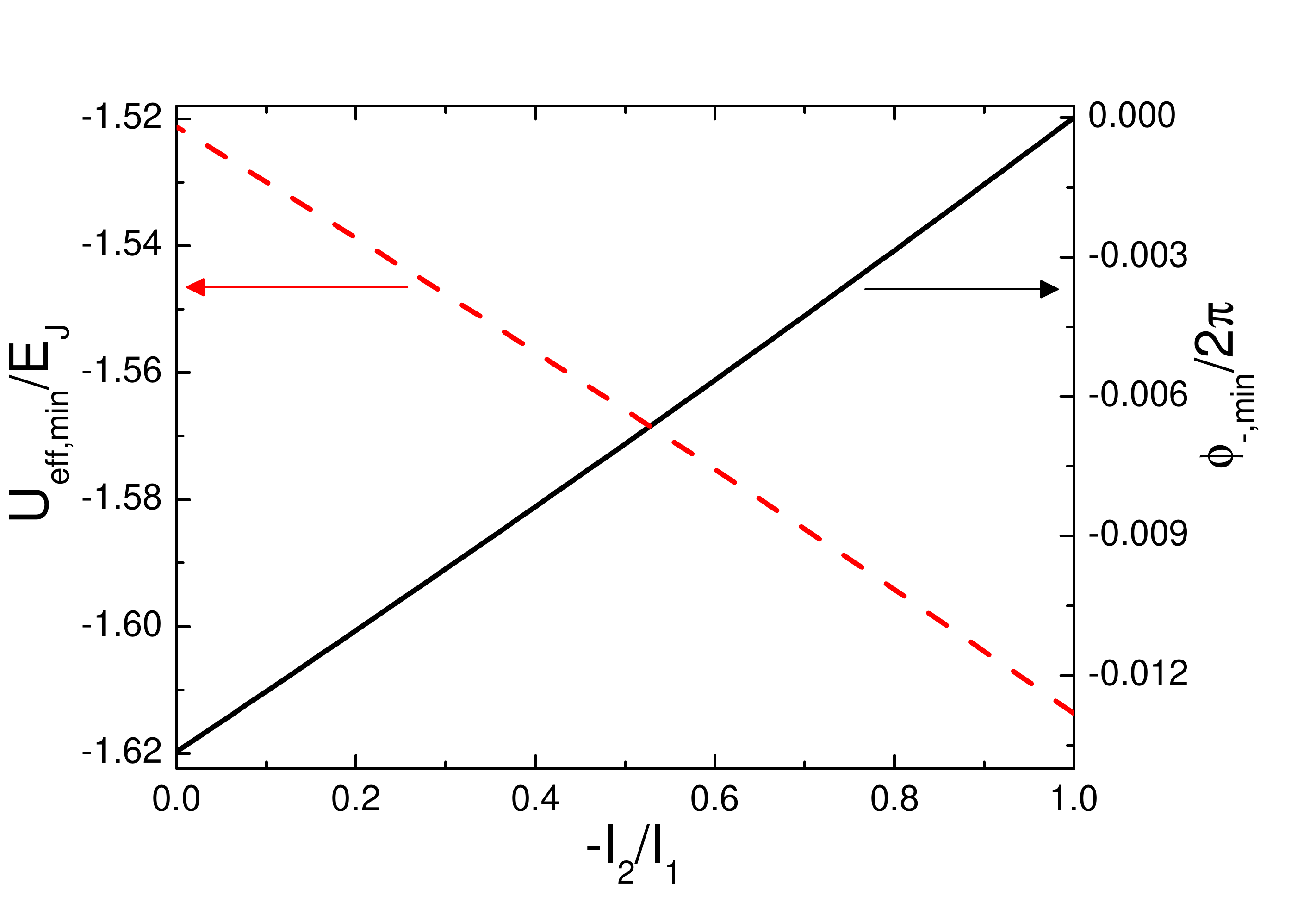}
\vspace{0cm}
\caption{The values of local minimum, $U_{\rm eff,min}$, of the effective potential
$U_{\rm eff}(\phi_+,\phi_-)$ around $\phi_+\approx \pi/3$ and $\phi_-\approx 0$ for $f=0.5$ (red dotted line)
and  $\phi_-$ at the local minimum, $\phi_{\rm -,min}$, (black solid line) are shown
as a function of $I_2/I_1$. }
\label{UminFig}
\end{figure}

The sum of first two terms, which corresponds to the effective potential of the flux qubit, can have local
minima for $\phi_-=0$ as shown in Fig. \ref{fig2}(a).
Here we set $I_1$ as the input bias current and
$I_2$ and $I_3$ as the output current so that $I_1=-I_2-I_3$.
The rms value of the resonator current is estimated as
${\cal I}_{\rm rms}=(1/\sqrt{2})\sqrt{\hbar\omega_{rj}/l_sL}=V_{\rm rms}/Z\sim $20nA,
where the impedance of the resonator $Z=50\Omega$ \cite{QIP} and the $V_{\rm rms}=E_{\rm rms}b=1\mu$V
with the electric field $E_{\rm rms}=0.2$V/m and the distance $b=5\mu$m \cite{Blais1}
between the transmission resonator and the ground plane.
Then the amplitude of bias current in Eq. (\ref{I0}) becomes
$|I_j|=2\sqrt{\hbar\omega_{rj}/l_sL}\sin(\pi d/L)\approx 14$pA,
where we set $L=25$mm and $d=2\mu$m.
Though the bias currents $I_j$ are much smaller than the critical current of
the flux qubit loop $I_c=2\pi E_{J1}/\Phi_0 \sim 500$nA, the finite bias
current may shift slightly the position of local minima of the total effective potential $U_{\rm eff}(\phi_+,\phi_-)$.

In Fig. \ref{UminFig} we show the numerical result for a local minimum, $U_{\rm eff,min}(I_2)$,  of
the effective potential $U_{\rm eff}(\phi_+,\phi_-)$  as a function of $I_2/I_1$, where
$\phi_{\rm +,min}\approx \pi/3$ and $\phi_{\rm -,min}\approx 0$
for $f=0.5$ at the local minimum. Here  the effective potential
minimum decreases until the output current $I_2$ reaches the maximum value, $-I_1$,
which can also be seen from the relation
$\partial U_{\rm eff,min}(I_2)/\partial I_2=(\Phi_0/2\pi)(\phi_{\rm +,min} +\phi_{\rm -,min})
\approx \Phi_0/6 > 0$  with the effective potential in Eq. (\ref{UeffT3}).
This means that the input current $I_1$ from the $j=1$ resonator flows through only
$j=2$ resonator ($I_2=-I_1$) while there is no output current flowing through the $j=3$
resonator ($I_3=0$).

As a result, at the local minimum the last term of the effective potential
$U_{\rm eff}(\phi_+,\phi_-)$ in Eq. (\ref{UeffT3})  reduces to
\begin{eqnarray}
\label{n3}
-\frac{\Phi_0}{2\pi}\phi_{+,{\rm min}}(I_1-I_2)
\end{eqnarray}
apart from the constant term independent of the phase variable $\phi_+$.
This term shifts the value of $\phi_{\rm +,min}$, but we still have $\phi_{\rm -,min}=0$ for $I_2=-I_1$
at the local minimum of $U_{\rm eff}(\phi_+,\phi_-)$ as  shown in Fig. \ref{UminFig}.
Here, we leave the term, $I_1-I_2$, in Eq. (\ref{n3}) which will be
represented in boson operators, $a_1$ and $a_2$, later.
The condition $\phi_{\rm -,min}=0$, i.e., $\phi_2=\phi_3$ means that the currents
$I'_i=-I_{ci}\sin\phi_i=-(2\pi E_{Ji}/\Phi_0)\sin\phi_i$ flowing across two junctions
are equal to each other, $I'_2=I'_3$, because $E_{J2}=E_{J3}$.
Thus from the current conservation $I'_2+I_3=I'_3$ we have $I_3$=0.
This is the reason why $I_3$ term in Eq. (\ref{UeffT3}) disappears in Eq. (\ref{n3}).
Therefore, by reducing one of the three Josephson junction energies
with threading flux $\Phi_{si}$ we can determine the output current channel,
which implements the current circulator function.
Here the sense of circulation can be determined {\it in situ} by choosing a dc-SQUID loop
to be threaded by the flux $\Phi_{si}$.

The main decoherence of flux qubit comes from the flux fluctuation, $\delta f$.
Though we consider $f=0.5$ in Fig. \ref{UminFig}, actually the relation
$\partial U_{\rm eff,min}(I_2)/\partial I_2 \approx \Phi_0/6 > 0$
is satisfied around $f\approx 0.5$. Hence, the value of $\phi_{\rm -,min}=0$ in Fig. \ref{UminFig}
is invariant and $I_2=-I_1$ at the local minimum
even in the presence of small flux fluctuations
so that the circulator function may be robust against the flux noise.


\begin{figure}[b]
\vspace{-0.5cm}
\centering
\includegraphics[width=14cm]{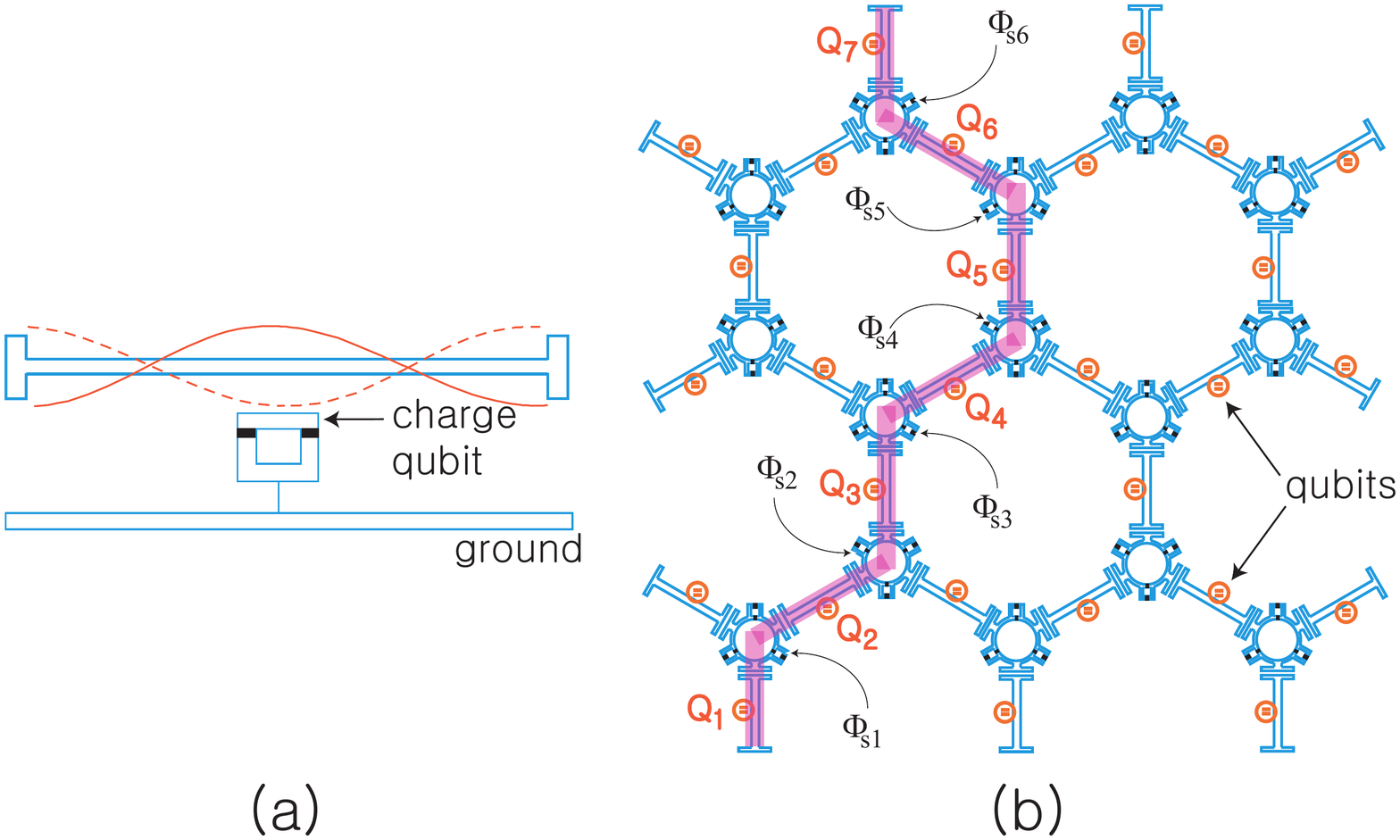}
\vspace{-1.5cm}
\caption{(a)Circuit-QED architecture with a qubit coupled to a transmission line resonator.
Here, we consider a superconducting charge qubit capacitively coupled to the resonator.
(b)Two-dimensional lattice based on the model
for coupling resonators in Fig. \ref{fig1}(b) for $n=3$, where
qubits coupled with the resonators form the Kagome lattice. Pair
of interacting qubits can be chosen by threading a flux
$\Phi_{si}$ into the dc-SQUID between the two corresponding
resonators.} \label{fig4}
\end{figure}

Similarly to the two-resonator case
the Hamiltonian in the tight-binding approximation
can be written with the effective potential in Eq. (\ref{UeffT3})
and the term in Eq. (\ref{n3}) as
\begin{eqnarray}
\label{H3}
H \!&=&\! E_{\downarrow}|\downarrow\rangle\langle\downarrow|\!+\!
E_{\uparrow}|\uparrow\rangle\langle\uparrow|
- t_q(|\downarrow\rangle\langle\uparrow|\!+\!|\uparrow\rangle\langle\downarrow|) \nonumber\\
&& -\frac{\Phi_0}{2\pi}\alpha(I_1-I_2) (|\downarrow\rangle\langle\downarrow| -
|\uparrow\rangle\langle\uparrow|),
\end{eqnarray}
where $\alpha=|\phi_{\rm +,min}|$.
By using Eq. (\ref{I0}) this Hamiltonian can be transformed to
\begin{eqnarray}
\label{Hc2} {{\cal H}}=\hbar\sum^3_{j=1}\omega_{rj}
a^\dagger_j a_j +\frac{1}{2}\omega_a\sigma_z+\frac{i}{2}
g\sigma_x[(a_1-a^\dagger_1)-(a_2-a^\dagger_2)]
\end{eqnarray}
with $g\approx \alpha \Phi_0\sqrt{\hbar\omega_{r}/l_sL}(d/L)$
in the basis of $|0\rangle$ and $|1\rangle$, where we set $\omega_{rj}=\omega_r$.
The last term of this Hamiltonian enables a selective interaction
between two resonators among three because there exits only the interaction
between the current modes of $j=1$ and $j=2$ resonators other than $j=3$.

If arbitrary one junction has a smaller Josephson junction energy
in Fig. \ref{fig1}(b) for $n=3$, the Hamiltonian is given by
\begin{eqnarray}
\label{Hc3}  {\tilde {\cal H}}=\hbar\sum^3_{j=1}\omega_{rj} a^\dagger_j a_j
+\frac{1}{2}\omega_a\sigma_z+\frac{i}{2}g\sigma_x[(a_l-a^\dagger_l)-(a_m-a^\dagger_m)],
\end{eqnarray}
where $(l,m)$=(1,2) if  the smaller Josephson junction energy  is $E_{J1}$.
If the smaller Josephson junction energy is $E_{J2}$ or $E_{J3}$,
$(l,m)$=(2,3) or (3,1), respectively. Here, the chirality
of the indices $(l,m)$ originates from the direction of the
threading external flux $\Phi_x$.
This Hamiltonian shows that only two resonators, $(l,m)$,
interact with each other while the other resonator does not
participate in the process.  Hence  one can decide the output channel
for a given input, and select two-resonator interaction.


For the general case of $n$ resonators connected at a vertex we consider that the $k$-th
Josephson junction energy is smaller than others such that
$E_{Jk}<E_{Ji}$ ($i\neq k$). With the flux quantization
condition $\phi_k=2\pi f-\sum^n_{i=1,i\neq k}\phi_i$ the condition
$\partial U_{\rm eff}(\{\phi_i\})/\partial \phi_i=0$ for extremal
point  results in $\phi_i=\phi$ for all $i ~(i\neq k)$ at the potential
minima. Then the last term of the effective potential $U_{\rm
eff}(\{\phi_i\})$ in Eq. (\ref{Ueffn}) becomes
\begin{eqnarray}
-\frac{\Phi_0}{4\pi}\phi\sum^n_{i=1}[n-1-2((n-i+k) ~{\rm mod}~ n)]I_i.
\end{eqnarray}
When $n=3$ and $k=1$, we can recover the result in Eq. (\ref{n3}) which describes the interaction
between two resonators. For $n>3$, however, this term contains more than two currents $I_i$
and thus we cannot obtain two resonator interaction any more. Hence in order to achieve interaction
between arbitrary two resonators connected to the same vertex we should consider $n=3$ case.

In order to describe the quantum gate operation between two qubits we
introduce a qubit coupled to a resonator as shown in Fig. \ref{fig4}(a). Here, we consider,
for example, a superconducting charge qubit capacitively coupled at the
center of the superconducting transmission line resonator through the second
harmonic voltage mode of resonator \cite{Blais}.
In Fig. \ref{fig4}(b) qubits $Q_i$ are introduced
at each resonator of the lattice for n = 3, where we can perform a quantum
gate operation between arbitrary two qubits among three qubits connected at a
vertex of a lattice.
These qubits interact with each other through the
resonator-resonator coupling. Among three qubits connected at a vertex arbitrary pair of two
qubits can interact with each other by threading a magnetic flux
$\Phi_{si}$ into the dc-SQUID between the two resonators coupled
with the qubits.

By extending the structure for $n=3$ we can have a
lattice structure in two-dimensional space as shown in Fig. \ref{fig4}(b).
The intervening flux qubits form the hexagonal lattice, but the qubits
coupled to the resonators form the Kagome lattice.
Two qubits $Q_1$ and $Q_2$, for example, can
interact with each other by threading the flux $\Phi_{s1}$,
and then two qubits $Q_2$ and $Q_3$ can
interact with each other by threading the flux $\Phi_{s2}$ after
switching off the coupling between qubits $Q_1$ and $Q_2$.
Switch-off can be done by threading a flux $\Phi_x$  far away from $0.5\Phi_0$
into the intervening flux qubit between qubits $Q_1$ and $Q_2$.
In this way arbitrary series of two-qubit gates with qubits, for example, $Q_1 \cdots
Q_7$ can  be performed in the Kagome lattice shown in Fig. \ref{fig4}.
As a result, we can achieve the effective interaction between two remote qubits, $Q_1$ and $Q_7$,
which is the key ingredient for the scalable quantum computing.

\section{Conclusion}

We have proposed a model for a scalable quantum computing in the Kagome
lattice based on the circuit-QED architecture. By controlling the flux threading
one of the three dc-SQUID loops of intervening flux qubit the circulator
function has been implemented in the superconducting circuit.
Hence we can control {\it in situ} the sense of circulation, which is a key ingredient in
the microwave quantum technology.
A scalable quantum computing requires
the selective interaction between two qubits among many qubits coupled at a
vertex point of the qubit lattice. In this study, we have shown that by using
the circulator function we can perform the selective two-qubit gate for the case
that only three qubits are coupled at the vertex through an intervening qubit,
where the qubit sites form the Kagome lattice. When more than three qubits
are coupled at a vertex, we have shown that the selective two-qubit coupling
cannot be achieved.

The two-qubit interaction between remote qubits are crucial for
the scalable quantum computing. In our design we can perform the
selective two-qubit gate between nearest neighbor qubits. Thus by
performing these two-qubit gates consecutively with switching
function we would achieve the quantum gate operation between
arbitrary pair of qubits in the lattice. We also have discussed
that the circulator function in this paper is robust against the
flux fluctuations, which will enable the experimental realization
of present scalable quantum computing model.

\begin{acknowledgements}
This work was partly supported by the KIST Institutional Program (Project No. 2E26680-16-P025)
and by Basic Science Research Program through the National Research Foundation
of Korea (NRF) funded by the Ministry of Education, Science and Technology (2011-0023467).
\end{acknowledgements}





\end{document}